\documentclass[11pt]{article}
\pdfoutput=1
\usepackage{graphicx}
\usepackage{amsmath,amsfonts,amssymb} 
\usepackage{color}
\usepackage{cite}
\usepackage{slashed}
\usepackage{hyperref}

\def\be{\begin{equation}}
\def\ee{\end{equation}}

\allowdisplaybreaks

\definecolor{myblue}{cmyk}{0.65, 0.37, 0.0, 0.19}
\begin{document}
\renewcommand*{\thefootnote}{\fnsymbol{footnote}}

\thispagestyle{empty}
\begin{flushright}
\rightline{BONN-TH-2018-05}

  \vspace*{2.mm} \today
\end{flushright}

\begin{center}
  {\Large \textbf{The non-Universal U(1) gauge extended $\mu\nu$SSM: anomalies cancellation and singular phenomenology
  } }  
  
  \vspace{0.5cm}
  V\'ictor Mart\'in-Lozano$^{1,}$\footnote{lozano$@$physik.uni-bonn.de},  Santiago Oviedo-Casado$^{2,}$\footnote{santiago.oviedo$@$upct.es} \\[0.2cm] 
    
  {\small \textit{ 
  $^1$Bethe Center for Theoretical Physics \& Physikalisches Institut der Universit\"{a}t Bonn,\\ Nu{\ss}allee 12, 53115, Bonn, Germany\\ 
          $^2$Departamento de F{\'i}sica Aplicada, Universidad Polit\'{e}cnica de Cartagena,\\ Cartagena 30202, Spain
   }}
  
\vspace*{0.7cm}

\begin{abstract}
So far the most sophisticated experiments have shown no trace of new physics at the TeV scale. Consequently, new models with unexplored parameter regions are necessary to explain current results, 
re-examine the existing data, and propose new experiments. In this Letter, we present a modified version of the $\mu\nu$SSM supersymmetric model where a non-Universal extra U(1) gauge symmetry is 
added in order to restore an effective R-parity that ensures proton stability. We show that anomalies equations cancel without having to add \emph{any} exotic matter, restricting the charges of the 
fields under the extra symmetry to a discrete set of values. We find that it is the viability of the model through anomalies cancellation what defines the conditions in which fermions interact with 
dark matter candidates via the exchange of $Z'$ bosons. The strict condition of universality violation  means that LHC constraints for a $Z'$ mass do not apply directly to our model, allowing for a 
yet undiscovered relatively light $Z$', as we discuss both in the phenomenological context and in its implications for possible flavour changing neutral currents. Moreover, we explore 
the possibility of isospin violating dark matter interactions; we observe that this interaction depends, surprisingly, on the Higgs charges under the new symmetry, both limiting the number of 
possible models and allowing to analyse indirect dark matter searches in the light of well defined, particular scenarios. 
\end{abstract}

\end{center}

\newpage
	
% -----------------------------------------------------------
% ARTICLE
% -----------------------------------------------------------

% INTRODUCTION
\renewcommand*{\thefootnote}{\arabic{footnote}}
\setcounter{footnote}{0}
\section{Introduction}
\label{sec:intro}

Barring the Higgs discovery \cite{HiggsATLAS,HiggsCMS}, no signs of new physics beyond the Standard Model (SM) have been seen so far after run 1 of the LHC. In particular, regarding Supersymmetry (SUSY), 
there are no signals of the coloured states \cite{GQATLAS}, namely squarks and gluinos, that were predicted to be abundant in the TeV scale. These data impose severe 
constraints on the allowed SUSY models, pushing the coloured states to masses beyond 1 TeV. However, several recast analyses showed that --even in already existing experimental data-- there still is much 
room for light SUSY states \cite{Arina:2016rbb, Kowalska2016,Buckley2017,Han2017,Kim:2017pvm}.

The minimal supersymmetric standard model (MSSM) is the most simple realisation of a $N=1$ SUSY model. In the MSSM construction however, the mass term responsible for the electroweak symmetry breaking 
(EWSB),i.e. the $\mu$-term, is added \textit{ad hoc}, not specifying its origin \cite{ABCSusy,TerningModernSusy,Drees,Susyprimer}. Another important issue is that the MSSM is unable to explain the fact that neutrinos do 
have mass \cite{Maltoni2004,Forero2014}. An elegant proposal to solve both problems at once comes from the so-called ``$\mu$ from $\nu$'' supersymmetric model ($\mu\nu$SSM)\footnote{See 
Ref.~\cite{Fidalgothesis} and references therein for an extensive review.}, which proposes introducing right-handed neutrinos to solve the $\mu$-problem and clarifying the origin of the left-handed neutrinos masses. The superpotential reads as follows,
\be
\begin{split}
W = &
\ \epsilon_{ab} \left(
Y_u^{ij} \, \hat H_2^b\, \hat Q^a_i \, \hat u_j^c +
Y_d^{ij} \, \hat H_1^a\, \hat Q^b_i \, \hat d_j^c +
Y_e^{ij} \, \hat H_1^a\, \hat L^b_i \, \hat e_j^c +
Y_\nu^{ij} \, \hat H_2^b\, \hat L^a_i \, \hat \nu^c_j 
\right) \\ &
-\epsilon_{ab} \lambda^{i} \, \hat \nu^c_i\,\hat H_1^a \hat H_2^b
+ \frac{1}{3}
 \kappa^{ijk} 
 \hat \nu^c_i\hat \nu^c_j\hat \nu^c_k\, .
\end{split} 
\label{superpotential1}
\ee
Two new terms in the superpotential of Eq.~\eqref{superpotential1} account for these properties: The first one, $\lambda_i \hat{\nu}_i^c \hat{H}_d \hat{H}_u$, $i=1,2,3$, is a trilinear 
coupling between the Higgs and the three families of right-handed neutrinos. When EWSB takes place the supersymmetric partners of the right-handed neutrinos, the right-handed sneutrinos 
$\widetilde{\nu}_{i}^c$, develop vacuum expectation values (VEVs) giving rise to an effective $\mu$ term, $\mu\sim\lambda v_{i}^c$. The second new term in the superpotential, 
$\kappa_{ijk}\hat{\nu}_i^c\hat{\nu}_j^c\hat{\nu}_k^c$, provides Majorana masses to the right-handed neutrinos after EWSB takes place. As the right-handed neutrinos couple to the leptons, a low-scale see-saw 
mechanism is induced and the light left-handed 
neutrinos become massive \cite{munuSSMoriginal,Fidalgo2009,Ghosh2008,Ghosh2010}.

As a result of the heavy mixing occurring within the neutral and charged sectors, the $\mu\nu$SSM presents a very rich phenomenology, markedly different from the usual collider scenarios \cite{Munoz2008,Ghosh2011,Ghosh2012,Ghosh2014,Munoz2016,Ghosh2017,Biekotter2017}. This means not only that new parameter regions open up for SUSY searches but also that the
$\mu\nu$SSM model predictions could have escaped unnoticed so far. Nonetheless the $\mu\nu$SSM has issues as well; both new terms added to the superpotential explicitly break $R$-parity ($R_p$) via the right-handed neutrinos, where such breaking is governed by the value of the neutrino Yukawa coupling, $Y_\nu$. As $R_p$ is no longer a symmetry of the model, dangerous lepton and 
baryon number violating terms are allowed in the superpotential. Likewise, the stability of the proton is no longer guaranteed. To recover an effective $R_p$ and at the same time allow only trilinear terms in the superpotential, one can invoke a U(1) gauge symmetry, which appears naturally in string realisations of the SM (see for example Ref.~\cite{FARAGGI2002,ibanez2014string,Marchesano2007,Shiu2005}).

The presence of an extra U(1) symmetry has already been explored both in the SM (see for example Ref.~\cite{Langacker2009,Martinez1,Martinez2}), as well as in supersymmetric realisations, of which 
Refs.~\cite{Loinaz1999,Ernest2002,Aoki1999,Aoki2000,Lee2008,Erler2000,Dobrescu1999,Demir2005,Japopesao} are only a few examples. In fact for the $\mu\nu$SSM it has already been tentatively explored 
 leading to promising results \cite{Fidalgo2012}. The price to pay however is having to recalculate the anomalies cancellation conditions, which for the SM matter content and gauge groups 
are known to be ``miraculously'' fulfilled. For example, in Ref.~\cite{Fidalgo2012} it was found that for the $\mu\nu$SSM to be consistent with an extra U(1) gauge symmetry, the matter content of the model has to be 
enlarged by several extra colour triplets, left doublets and singlet fields. Moreover, it is a general rule for all gauge extended models that exotic fields are needed for the model to be anomaly free. While the possibility of exotic fields cannot be excluded, its presence is problematic, not only due to the lack of evidence but because it disrupts the unification of coupling constants at the GUT scale. Hence a minimalist solution is always desired. 

In this Letter we present a solution of the U(1) enlarged $\mu\nu$SSM which is anomaly free by means of having non-Universal charges of the superfields under the extra symmetry, with the novelty that no exotic fields are needed. We solve the anomalies equations by assuming that each family can have a 
different U(1)$_X$ charge, finding several groups of solutions depending on few mostly unconstrained extra charges. In addition, we explore some possible phenomenological consequences of having non-Universality in our model. 
Concretely, we study the extra charge dependence of the $Z'$ interaction with fermions as a result of the mixing between the extra U(1) boson and the usual $Z$ boson \cite{Soler2014,Feng2014}. Such 
dependence implies that the production limits of a $Z'$ at the LHC are no longer valid and have to be recalculated for the specific models allowed by the anomalies cancellation, leading to scenarios 
where a light $Z'$ is possible, a common feature of string constructions \cite{Soler2014,Ringwald2009,Quevedo2002}, but which is however bounded from below by the condition that no flavour 
changing neutral currents (FCNC) --common in models with extra symmetries \cite{Langacker2000,Langacker2004}-- have been found experimentally \cite{ATLASfcnc,CMSfcnc}, as we explain. Furthermore, we 
explore possible scenarios of $Z'$ mediated spin independent dark matter (DM) interactions \cite{Chun2011}, finding a family of anomalies equations solutions with both scalar and vector isospin 
violating DM-quark interactions. For the non-Universal U(1) gauge enlarged $\mu\nu$SSM the possible isospin violating scenarios are parametrised by the Higgs fields charges under the extra symmetry, 
rendering a series of finite, discrete values that could be discriminated in experiments. Therefore in our model experimental detection of DM is not only conditioned by the specific realisation but 
could also be used to provide with clear, testable predictions to discern among DM and string compactification scenarios.

%MODEL

\section{The non-Universal U(1)$\mu\nu$SSM}

In this section we present the necessary conditions for anomalies cancellations and the implications for model building. In addition, we explore possible phenomenological implications and signatures particular of the model that are imposed by the extra charges assignment of the fields, which themselves are constrained by the anomalies cancellation conditions.

\subsection{Non-universal anomaly cancellation in the U(1)$\mu\nu$SSM}

The gauge group of the U(1)$\mu\nu$SSM is $SU(3)_C \times SU(2)_L \times U(1)_Y \times U(1)_X$, where each of the superfields composing the model's spectrum has now an extra,$\,Q_X\,$, 
charge. Consequently, all anomalies equations involving the new symmetry have to be recalculated if we want the model to be anomaly free. The analysis is two-fold: On the one hand, the terms 
appearing in the superpotential must have vanishing total charge, and on the other hand, anomalies must also cancel. Hence, restrictive bounds are imposed on the possible values that each superfield charge $Q_X$ can have. Furthermore, we can use certain constraints to either allow certain terms in the superpotential, or to explicitly banish undesired, unphysical, or dangerous terms.

We will work under the assumption that no exotic matter is needed. To that end, we consider the same matter content as for the original $\mu\nu$SSM. The anomalies equations that must be fulfilled with 
this matter content are
\be
\begin{split} 
 & \sum_i(2Q_{Q_i}+Q_{u_i}+Q_{d_i}) = 0, \\ &
 \sum_i(3Q_{Q_i}+Q_{L_i})+Q_{H_1}+Q_{H_2} = 0, \\ &
\sum_i(\frac{1}{6}Q_{Q_i}+\frac{1}{3}Q_{d_i}+
 \frac{4}{3}Q_{u_i}+\frac{1}{2}Q_{L_i}+Q_{e_i})+
 \frac{1}{2}(Q_{H_1}+Q_{H_2}) = 0, \\ &
\sum_i(Q_{Q_i}^2+Q_{d_i}^2-2Q_{u_i}^2-Q_{L_i}^2+Q_{e_i}^2)-
 Q_{H_1}^2+Q_{H_2}^2 = 0, \\ &
\sum_i(6Q_{Q_i}^3+3Q_{d_i}^3+3Q_{u_i}^3+2Q_{L_i}^3+Q_{e_i}^3 + Q_{\nu^c_i}^3)+
 2Q_{H_1}^3+2Q_{H_2}^3 = 0, \\ &
\sum_i(6Q_{Q_i}+3Q_{u_i}+3Q_{d_i}+2Q_{L_i}+Q_{e_i}+Q_{\nu^c_i})+2Q_{H_1} 
 +2Q_{H_2} = 0.
\end{split}
\label{anomalies}
\ee
To solve equations from Eq.~\eqref{anomalies} we need a set of constraints, which we in addition use to ensure that our model has certain desired properties arising naturally. Prime among them is forbidding 
a bilinear $\mu$ term from appearing in the superpotential, as its absence is otherwise not automatically guaranteed. Thus, we impose $Q_{H_1} \ne - Q_{H_2}$. Furthermore, since the Higgs mass term is obtained from the right-handed sneutrinos $\nu^c$ singlet fields acquiring VEVs at the EWSB scale, a term coupling the right-handed neutrinos and the Higgs fields must also be allowed in the superpotential, which requires $Q_{H_1} + Q_{H_2} + Q_{\nu^c_i} = 0$ for at least one of the three families of right-handed neutrinos. 
Moreover, as the $\mu\nu$SSM was born to answer the neutrino mass problem, and the extra U(1) forbids the presence of the $\kappa_{ijk}\hat{\nu}_i^c\hat{\nu}_j^c\hat{\nu}_k^c$ term that provided Majorana masses for the right-handed neutrinos in the original $\mu\nu$SSM, we impose that Yukawa tree-level mass terms below the soft breaking scale must appear for the right-handed neutrinos, such that a see-saw mechanism is implemented in our model. It is therefore a 
condition that $Q_{l_i} + Q_{\nu_i^c} + Q_{H_2}$ = 0. In addition, we would like to directly forbid certain operators --such as those violating baryon number-- from the superpotential, which means 
$Q_u \ne -2Q_d$. The remaining mass terms are \emph{a priori} not imposed in the superpotential, permitting thus the different fields to acquire mass either at tree-level order with Yukawa couplings 
or at first loop, via non-holomorphic mass terms. The choice of either is to be fixed accordingly with the anomalies equations.

Giving mass to certain fields via non-holomorphic terms means that such mass must be provided by SUSY-breaking operators introduced via radiative, first loop corrections, which appear 
naturally in gravity mediated SUSY-breaking scenarios \cite{Martin2000}. It has been demonstrated that either mechanism is in principle indistinguishable in experiments but for the heaviest particles 
\cite{BORZUMATI199953}, namely the top quark and the $\tau$ lepton, for which tree level Yukawa terms need to be imposed. Thus, we can impose $Q_{Q_3} + Q_{u_3} + Q_{H_2}$ = 0 
and $Q_{l_3} + Q_{e_3} 
+ Q_{H_1}$ = 0. With this conditions, the first, second, and sixth equations in Eq.~\eqref{anomalies} become 
\be
\begin{split}
&2Q_{Q_1}+2Q_{Q_2}+Q_{Q_3}+Q_{d_1}+Q_{d_2}+Q_{d_3}+Q_{u_1}+Q_{u_2}-Q_{H_2} = 0, \\&
Q_{H_1} + Q_{H_2} + Q_{L_1} + Q_{L_2} + Q_{L_3} + 3Q_{Q_1} + 3Q_{Q_2} + 3Q_{Q_3} = 0, \\&
3Q_{d_1} + 3Q_{d_2} + 3Q_{d_3} + Q_{e_1} + Q_{e_2} + Q_{H_1} + Q_{L_1} + Q_{L_2}\\& + 6Q_{Q_1} + 6Q_{Q_2} + 3Q_{Q_3} +  3Q_{u_1} + 3Q_{u_1} - 4Q_{H_2} = 0,
\end{split}
\ee
which we can use to fix the conditions for quark charges. Should we try for all the up quarks to have tree-level mass terms, then all down quarks necessarily acquire mass through non-holomorphic terms. But 
from the third equation this imposes that just one lepton has tree-level Yukawa mass term. Putting everything back into the second equation it would lead to $Q_{H_1} = - Q_{H_2}$, reintroducing the
$\mu$ term in the superpotential. And the same happens if we try to have both first and second families of up quarks with non-holomorphic mass terms. Consequently, the only possibility is for either the 
first or the second family of up quarks to have tree-level Yukawa coupling, while the other acquires its mass through a non-holomorphic term. On the contrary, the necessary condition for the down-type quarks is 
to have \emph{two} of the families having non-holomorphic mass terms and one a tree-level Yukawa. There is however freedom in choosing which family acquires its mass via which mechanism, a fact that will be of importance for the phenomenology of the model as we shall see in what follows. The left leptons (L) mimic the behaviour of the up-quarks. The possibility of flavour changing neutral currents in both quark and lepton sectors can as well be disregarded as the non-diagonal flavour matrices terms can be safely made zero. Furthermore, even if the mechanism for obtaining the masses changes within the quark (or lepton) type, no problem exists in obtaining the correct masses for each of the particles.

To conclude the analysis, we must choose either of the groups of solutions, the rest being symmetric. In particular, fixing the first and third families of up-quarks with superpotential tree-level Yukawa 
couplings, and establishing the remaining quarks and leptons accordingly, immediately fulfils equations 1 and 6 from  Eq.~\eqref{anomalies}, and leaves the second and third the same and equal to
\be
Q_{H_1} + Q_{H_2} + Q_{L_1} + Q_{L_2} + Q_{L_3} + 3Q_{Q_1} + 3Q_{Q_2} + 3Q_{Q_3} = 0.
\ee
Clearing $Q_{L_1}$ and replacing it in the non-linear anomalies we obtain for equation four in Eq.~\eqref{anomalies}
\be
(Q_{H_1} + Q_{H_2}) (Q_{H_1} + Q_{L_2} + 3 (Q_{Q_1} + Q_{Q_3})) = 0.
\ee
Choosing the first brackets to hold true would reintroduce the $\mu$ term in the superpotential, therefore is the second brackets what must cancel, fixing $Q_{L_2} = -Q_{H_1} - 3Q_{Q_1} - 3 Q_{Q_3}$, 
with which the fifth equation in Eq.~\eqref{anomalies} is simplified to 
\be
\begin{split}
&(Q_{H_1} + Q_{H_2}) \\&(-2Q_{L_3} (Q_{H_2} +Q_{L_3}) + Q_{H_1}(Q_{H_2} - 3Q_{Q_2}) - 3 (Q_{H_2} + 2Q_{L_3})Q_{Q_2} - 9Q_{Q_2}^2) = 0.
\end{split}
\label{lastineq}
\ee

We thus have solutions depending on $Q_{H_1}$,$Q_{H_2}$,$Q_{L_3}$,$Q_{Q_1}$,$Q_{Q_2}$,$Q_{Q_3}$, a repeating characteristic of the model independently of which fields have non-holomorphic mass term, 
and where any combination is valid as long as Eq.~\eqref{lastineq} has a real solution and a term of the form $Q_{H_1} + Q_{H_2} + Q_{\nu^c_i} = 0$ is allowed. For this to happen, the corresponding 
$Q_{L_i}$ must be equal to $Q_{H_1}$, which means that not all right-handed neutrinos will have a tree-level coupling with the Higgs fields, being nonetheless guaranteed that some will, thus providing a natural mass 
term for the Higgs particle. In addition, note that by having the quark families mass terms with opposite mechanisms, it is guaranteed that no baryon number violating operator is allowed in the superpotential as 
long as Higgs and quarks have different charges. We thus have shown that within the framework of the U(1) extended $\mu\nu$SSM, anomalies are cancelled without the need to add exotic matter, at the 
price of having non-universal charges, and with the gain of forbidding most troublesome operators in the superpotential. In the appendix \ref{sec:app}, an specific $Q_{X}$ charge distribution for the above described family of 
solutions can be found, together with and an altogether different scenario with important DM phenomenological consequences. In the remaining part of the article we elaborate on the novel, particular, 
and potentially relevant phenomenological characteristics of the non-Universal U(1)$\mu\nu$SSM.

\subsection{Phenomenological consequences of Universality violation across fermion families.}

The phenomenological manifestation of an extra U(1) gauge symmetry comes, mainly, from the mixing of the massive neutral components of the vector bosons from the gauge sector, namely the $Z$ and the $Z'$ gauge bosons. For the case of the U(1) enlarged $\mu\nu$SSM, and contrary to other models where similar mixing occurs (see for example Ref.~\cite{Bandyopadhyay:2018cwu} and references therein), the mixing happens naturally within the neutralino mixing matrix as part of the right sneutrinos acquiring vacuum expectation value. Such fact enriches greatly the phenomenology of the model and, as we will describe, imposes conditions for both collider and dark matter interactions, similar to what happens within the $\mu\nu$SSM alone \cite{Fidalgothesis}. The presence of the non-Universal extra gauge symmetry introduces a new dependence on the specific charge of each field under the extra symmetry which will condition the possible interactions of fermions with the $Z'$, which is responsible of a very particular phenomenology, specific of the non-Universal U(1)$\mu\nu$SSM. 

The phenomenology of the model is modified according to the the mixing of the new sector. To parametrise the influence that the extra U(1) has, we define the mixing factor $R$ \cite{Fidalgo2012}, 
\begin{equation}
R=\frac{(M_{ZZ'}^2)^2}{M^2_Z M^2_{Z'}},
\end{equation}
where the entries $M_{ij}$ correspond to the terms of the mixing matrix between $Z$ and $Z'$. In principle $R$ should be smaller than $10^{-3}$ given the experimental constraints available 
\cite{Cho:2000pq}, with the consequence that $M_{ZZ'}$ has to be smaller than $M_{ZZ'} \lesssim 56$ GeV \footnote{A complete description of the entries $M_{ij}$ and their dependencies with the parameter of the model can be found in Ref.~\cite{Fidalgo2012}.} when $M_{Z'}$=1 TeV, with such constraint becoming weaker as the mass of the $Z'$ gets heavier. Thus only heavier masses for the $Z'$ would fulfil such condition together with the ones coming from accelerator searches, and would require a somewhat large fine-tuning. Nonetheless, these limits are calculated for when the extra charges are 
Universal, which does not occur in our model. Hence the bounds presented have to be taken carefully, as the couplings of the physical states are now dependent on the $Q_X$ charges (as well as on the vacuum 
expectation value of $\nu^c$). However, in the rest of the paper we will consider that the mixing in the $Z-Z'$ sector is negligible.\footnote{ Notice nonetheless the new configuration of charges and the relation among the different vacuum expectation values can induce a sizeable mixing in the $Z-Z'$ system.}

In our model the physical coupling of the $Z'$ to the fermionic sector is 
\begin{equation}
g_\alpha^{Z'}=g' Q_X^{\alpha},
\label{eq:charge}
\end{equation}
where $\alpha$ corresponds to the matter field $\psi_\alpha$, $Q_X^\alpha$ is the charge of this field under the U(1)$_X$, and $g'$ corresponds to the coupling constant of the U(1)$_X$ gauge  
symmetry. Thus, once  the value of $g'$ is fixed, the way the $Z'$ couples to the different fermions depends strictly on the charges $Q_X$. The values these charges can have are fixed by the   
anomalies cancellation conditions, with only certain discrete values allowed. Moreover, since these charges break universality among fermions (see for example the models presented in the    
Appendix~\ref{sec:app}), each fermion family will have in general a different value of the charge and consequently will couple with different strength to the $Z'$, having deep phenomenological consequences.

The physical couplings of the SM particle to the $Z'$ are described as follows. According to Eq.~\eqref{eq:charge} the left and right handed components of the SM fermions do not necessarily      
share the same couplings to the $Z'$, as they depend on the charge assignation. Usually, the couplings of a vector boson can be expressed in its vector and axial forms. The vector coupling is  
defined  as the sum of the left and right components, for example the vector coupling of the quarks is,
\begin{eqnarray}
C_{u_i}^V =  g_{u_{iL}}^{Z'} + g_{u_{iR}}^{Z'} = g' (Q_{u_{iL}}+Q_{u_{iR}}), \\
C_{d_i}^V =  g_{d_{iL}}^{Z'} + g_{d_{iR}}^{Z'} = g' (Q_{d_{iL}}+Q_{d_{iR}}),
\label{veccoupling}
\end{eqnarray}
where $i=1,2,3$ stands for the three families of both up and down quarks since they could have different charges. On the contrary, the axial coupling is defined as the difference of the          
components,
\begin{eqnarray}
C_{u_i}^A =  g_{u_{iL}}^{Z'} - g_{u_{iR}}^{Z'} = g' (Q_{u_{iL}}-Q_{u_{iR}}), \\
C_{d_i}^A =  g_{d_{iL}}^{Z'} - g_{d_{iR}}^{Z'} = g' (Q_{d_{iL}}-Q_{d_{iR}}),
\end{eqnarray}
As we see, the vector and axial couplings are different, which should not surprise us since in the SM the $Z$ boson behaves similarly. As the charges can be different in the up and down sectors the 
result is that $Z'$ does not couple in the same way to up and down quarks. The consequences are twofold. On the one hand, the production rates of a $Z'$ in collider experiments has to be recalculated 
taking into account that each of the fermion pairs that could produce a $Z'$ has a different value of the coupling, which makes the current constraints and limits invalid in this model. And on the other hand, direct dark 
matter searches, which are heavily dependent on the DM particle interaction with protons and neutrons, are affected by the fact that now the coupling is quark-family dependent, and will not   
interact the same with protons (which have more up-quark content) than with neutrons, modifying as well current experimental DM searches and imposing different limits.

In the remaining part of the section we will describe briefly the interesting phenomenological consequences that we just commented, focusing on the collider and DM.\footnote{A more 
detailed and involved description of the phenomenology of these models will be described in a forthcoming work~\cite{nextone}}

%%%%%%%%%%%%%%%%%%%%%%%%%%%%%%%%%%%%%%%%%%%%%%%%%%%%%%%%%%%%%%%%%%%%%%%%%%%%%%%%%%%%%%%%%%%%%%%%%%%%%%%%%%%%%%%%%%%%%%%%%%%%%%%%%%%%%%%%%%%%%%%%%%%%%%%%%%%%%%%%%%%%%%%%%%%%%%%%%%%%%%%%%%%%%%%%%%%%%%%%%%%%%%%%

\vspace*{0.5cm}

\noindent\textbf{On the existence of unobserved flavour changing neutral currents}

\vspace*{0.5cm}

The presence of flavour changing neutral currents is highly suppressed in the SM \cite{ATLASfcnc,CMSfcnc}. In that sense, non-Universality can be problematic, as is the distribution of extra 
charges what determines the Yukawa textures, which in turn can lead to differences in the CKM matrix for quarks and in the lepton currents, introducing new $Z$' mediated FCNC. To understand 
the mechanism by which the extra U(1) symmetry might introduce FCNC, it is illustrative to write the Q's extra charges matrices associated to the Yukawa mass operators. In particular, for 
up-squarks, both the holomorphic ($Y_u^{ij} \, \hat H_2^b\, \hat Q^a_i \, \hat u_j^c$) and non-holomorphic ($Y_u^{ij} \, \hat H_1^b\, \hat Q^a_i \, \hat u_j^c$) charge matrices look like,
\be
Q_{Q_i} + Q_{u_j} + Q_{H_2} = \left(
\begin{array}{ccc}
0 & Q_{Q_1} - Q_{Q_2} + Q_{H_1} + Q_{H_2} & Q_{Q_1} - Q_{Q_3} \\
Q_{Q_2} - Q_{Q_1}  &  Q_{H_1} + Q_{H_2} & Q_{Q_2} - Q_{Q_3} \\
Q_{Q_3} - Q_{Q_1} & Q_{Q_3} - Q_{Q_2} + Q_{H_1} + Q_{H_2} & 0 
\end{array}
\right),
\label{quarks1}
\ee
\be
Q_{Q_i} + Q_{u_j} - Q_{H_1} = \left(
\begin{array}{ccc}
-Q_{H_1} - Q_{H_2} & Q_{Q_1} - Q_{Q_2}  & Q_{Q_1} - Q_{Q_3} -Q_{H_1} - Q_{H_2} \\
Q_{Q_2} - Q_{Q_1} -Q_{H_1} - Q_{H_2}  &  0 & Q_{Q_2} - Q_{Q_3} -Q_{H_1} - Q_{H_2} \\
Q_{Q_3} - Q_{Q_1} -Q_{H_1} - Q_{H_2} & Q_{Q_3} - Q_{Q_2} & -Q_{H_1} - Q_{H_2} 
\end{array}
\right).
\label{quarks2}
\ee

The case of down squarks is symmetric to the up squarks, interchanging the tree level and non-holomorphic matrices, while for right sleptons the behaviour is the same as for 
up squarks (replacing the corresponding $Q_Q$ by $Q_L$, while for right sneutrinos, since we choose them to all have tree level Yukawas, no problems arise with FCNCs. What Eqs.~\eqref{quarks1} and 
\eqref{quarks2} tell us is that in order for a Yukawa term to be allowed in the superpotential, the corresponding matrix entry must be zero. Hence, for a FCNC to appear in the superpotential, 
non-diagonal entries in Eqs.~\eqref{quarks1} or \eqref{quarks2}, and their peers for down squarks and sleptons must be zero. Thus, for sleptons avoiding unwanted, $Z'$ mediated FCNCs is easy, as it only 
requires flavour changing terms to be forbidden in the Yukawa matrices, which will occur as long as $Q_{L_i} \ne Q_{L_j}$ with $ i \ne j$, since it is already guaranteed that the Higgs charges have to 
be different from one another. Consequently no $Z'$ mediated leptons FCNC will appear in this model. For the remaining we will thence concentrate on the possibility of FCNC in the quark sector. 

$Z'$ mediated quark neutral currents are governed by the coupling of the $Z'$ to the quarks, which we assume to be diagonal in the weak basis. This is easily achieved as long as the non-diagonal terms 
in Eqs.~\eqref{quarks1} and \eqref{quarks2} are different from zero, as it happens with sleptons. In whose case the Lagrangian looks like
\be
\mathcal{L} = -\frac{g}{2\cos\theta_w}\left( \tilde{U}_L\delta^U_L\gamma_\mu U_L + \tilde{U}_R\delta^U_R\gamma_\mu U_R + \tilde{D}_L\delta^D_L\gamma_\mu D_L + \tilde{D}_R\delta^D_R\gamma_\mu D_R \right)Z^{'\mu}.
\ee
Here $\tan\theta_w = g_Y/g'$, the ratio between the hypercharge and the new U(1) coupling constants, $U_{L,R} = (u,c,t)^T_{L,R}$, $D_{L,R} = (d,s,b)^T_{L,R}$, and $\delta^{U,D}_{L,R}$ is the Kronecker delta for left-right and up-down type terms, indicating that 
there are no non-diagonal couplings in the Lagrangian. Therefore, possible interactions which are flavour changing will come described by the rotation of the Yukawa quark matrices to the physical 
bases, i.e. the mass eigenstates. If these rotations were proportional to the unit matrix, then no FCNC would appear. As it stands, there is no guarantee that the interaction remains diagonal, hence 
the corresponding Lagrangian is \cite{Valencia2002,Valencia2009}
\be
\begin{split}
\mathcal{L}_{FCNC} &= -\frac{g}{2\cos\theta_w} ( \tilde{U}_L\gamma^\mu V^U_L \delta^U_L {V^U_L}^\dagger U_L + \tilde{U}_R\gamma^\mu V^U_R \delta^U_R {V^U_R}^\dagger U_R \\& + \tilde{D}_L\gamma^\mu 
V^D_L \delta^D_L 
{V^D_L}^\dagger D_L + \tilde{D}_R\gamma^\mu V^D_R \delta^D_R {V^D_R}^\dagger D_R )Z'^{\mu}, 
\end{split}
\ee
where the $V_{L,R}^{U,D}$ are the usual quark diagonalisation matrices, the well known Kobayashi-Maskawa. The flavour changing neutral currents $J_{Z'}$ associated to $\mathcal{L}_{FCNC}$ in terms of the Kobayashi-Maskawa mixing terms are of the form~\cite{Valencia2002},
\be
\sin\theta_w\cot\theta_X\cos\xi_Z V^{U,D}_{L,R}{V^{U,D}_{L,R}}^\dagger,
\ee
with $\tan\theta_X = g/g'$, that is the ratio of the weak and the U(1) gauge coupling constants. $\tan 2\xi_Z = 2 M_{ZZ'}^2/(M_{Z}^2 + M_{Z'}^2)$, where the entries $M_{ij}$ correspond to the terms of the mixing matrix between $Z$ and $Z'$. 

Then whether FCNC appear and are important in our model is a matter of diagonalizing the Yukawa quark matrices for specific realisations of the extra charges and obtaining the corresponding 
neutral 
currents and their specific strength. The constraint for their appearance goes like \cite{Valencia2013}
\be
\frac{M_Z}{M_Z'}\sin\theta_w\cot\theta_X\cos\xi_Z V^{U,D}_{L,R}{V^{U,D}_{L,R}}^\dagger \lessapprox 10^{-4},
\label{eqguay}
\ee
which for typical parameters means $\frac{M_Z}{M_Z'} \lessapprox 1$.

Given the fact that the mixing between the states is almost negligible the contribution to the FCNC through the mixing will be suppressed. Apart from that, a $Z'$ with a mass in the TeV range has 
practically a negligible effect on FCNC, as can be deduced from Eq.~\eqref{eqguay} \cite{Langacker:2008yv,Valencia2013}. The LHC searches for a $Z'$ set the mass of this boson to be in the multi-TeV range 
in order not to be produced. In order to be safe from such constraints, one should invoke either a large TeV mass or a really small $U(1)'$ gauge coupling. In both scenarios, the total contribution 
of the $Z'$ is practically negligible since the contributions are suppressed either to a high $m_{Z'}$ or a really small coupling $g'$. As the scope of this paper is not a detailed and numerical study 
of the properties of this model but just a broad overview of the interesting phenomenological aspects, we leave the precision calculations and numerical results to a future and deeper study of this 
model~\cite{nextone}.

%%%%%%%%%%%%%%%%%%%%%%%%%%%%%%%%%%%%%%%%%%%%%%%%%%%%%%%%%%%%%%%%%%%%%%%%%%%%%%%%%%%%%%%%%%%%%%%%%%%%%%%%%%%%%%%%%%%%%%%%%%%%%%%%%%%%%%%%%%%%%%%%%%%%%%%%%%%%%%%%%%%%%%%%%%%%%%%%%%%%%%%%%%%%%%%%%%%%%%%%%%%%%%%%

\vspace*{0.5cm}

\noindent\textbf{Collider Phenomenology of the $Z'$}

\vspace*{0.5cm}

The $Z'$ could be eventually produced in the LHC. Both ATLAS and CMS have searches on high mass resonances decaying into a pair of leptons or hadronically (see for example Ref.~ \cite{Aaboud:2017buh,Aaboud:2017sjh,Sirunyan:2018exx}). As no signal of a $Z'$ has yet been found, bounds can be set on the production and subsequent decay of a $Z'$, $pp\rightarrow Z' \rightarrow \psi\bar{\psi}$ for a defined mass. However, in the set of non-Universal models one can avoid such strong limits provided that different fermion families couple differently to the $Z'$. It could be the case that the up and down quark families have charges such that the effective coupling to the $Z'$ gets suppressed together with its production.

The general expression for the $Z'$ production and subsequent decay into fermions at the LHC is~\cite{Martin-Lozano:2015vva, Bandyopadhyay:2018cwu},
\begin{equation}
\sigma_{f\bar{f}}\simeq \left(\frac{1}{3}\sum_q\frac{dL_{q\bar{q}}}{dm_{Z'}^2}\times \hat{\sigma}(q\bar{q}\rightarrow Z')  \right) \times BR(Z'\rightarrow f\bar{f}),
\end{equation}
where $dL_{q\bar{q}/dm_{Z'}^2}$ stands for the parton luminosities, $\hat{\sigma}(q\bar{q}\rightarrow Z')$ is the peak cross section for the $Z'$ boson, and $BR(Z'\rightarrow f\bar{f})$ is the branching ratio for the $Z'$ decaying into a fermion pair. As it was pointed out in Ref.~\cite{Martin-Lozano:2015vva}, one can describe those parameters as a function of the sum of the different production rates for each quark and its $Z'$ coupling,
\begin{equation}
\sigma_{f\bar{f}}^{\rm LO}=\sum_{i=1}^{3}\left[c_{u_{i}}\tilde{\omega}_{u_{i}}(s,m_{Z'}^2)+c_{d_{i}}\tilde{\omega}_{d_{i}}(s,m_{Z'}^2)\right] \times {\rm BR}(Z'\rightarrow f\bar{f}).
\end{equation}
Here, $c_{q}$ are defined as $c_{q}= (C^V_{q})^2 + (C^A_{q})^2$ and the functions $\tilde{\omega}_q(s,m_{Z'}^2)$ contain all the information related with the parton distribution function, NLO corrections, etc.\footnote{For further information one can see Ref.~\cite{Martin-Lozano:2015vva} or the Appendix of Ref.~\cite{Bandyopadhyay:2018cwu}}

The most important part of the model in the $Z'$ collider phenomenology is the fact that all type of fermions, no matter the family or the flavour, couple differently to the $Z'$. This weakens the experimental searches of this kind of particles that ATLAS and CMS perform. The are several ways in which the $Z'$ production might be diminished. One can have small quark couplings giving a tiny production cross section in such a way that the $Z'$ is barely produced in the LHC even for light masses of the $Z'$. Together with this effect the charges to the leptonic sector could be small as well reducing the total amount of observable events. However, in this model the charge assignment is not free since it is fixed by the cancellation of anomalies. As a consequence, one cannot arbitrarily make the couplings as small as it would be required to directly avoid collider searches, and the specific model realisation completely determines the $Z'$ phenomenology, which means that clear, precise predictions for the LHC can be established; on the other hand, a $Z'$ discovery would severely constraint the possible models, thus hinting towards the specific realisation in nature of the non-Universal U(1)$\mu\nu$SSM. In that sense a deeper study will be done in the future~\cite{nextone} to determine the consequences of such charge assignment. 

\vspace*{0.5cm}

\noindent\textbf{Dark Matter Phenomenology }

\vspace*{0.5cm}

There are different candidates for a dark matter particle in the non-Universal U(1)$\mu\nu$SSM. Among them, the role could be played by an extra vector-doublet, inert in the SM sector, a decoupled field, such a non-interacting right-handed neutrino, or as it occurs in the $\mu\nu$SSM, the gravitino \cite{Fidalgothesis}. The interesting scenario occurs in the first two cases, where a DM distinctive signal comes from the $Z'$ mediated spin independent interaction with a dark matter particle $\psi$. We can parametrise the effective Lagrangian of DM particle interaction with protons $p$, and neutrons, $n$, mediated by a vector boson as,
\begin{equation}
\mathcal{L}^{\rm SI}_{\rm V}= f_p(\bar{\psi}\gamma_\mu \psi)(\bar{p}\gamma^\mu p) + f_n (\bar{\psi}\gamma_\mu \psi)(\bar{n}\gamma^\mu n),
\end{equation}
where the vector couplings $f_p$ and $f_n$ are defined through their nucleon valence quark content as \cite{Chun2011}
\begin{equation}
f_p=2b_u + b_d, f_n= b_u + 2b_d,
\end{equation}
with $b_{u,d}$ the effective $Z'$ mediated vector couplings 
\begin{equation}
b_{(u,d)}=\frac{g_{dm}C^V_{(u,d)}}{2m_{Z'}^2}.
\end{equation}
Using the definitions for the vector coupling obtained in Eq.~\eqref{veccoupling} we have that
\begin{equation}
b_{(u)}=\frac{g_{dm}g'}{2m_{Z'}^2}(Q_{u_L}+Q_{u_R}),
\end{equation}
\begin{equation}
b_{(d)}=\frac{g_{dm}g'}{2m_{Z'}^2}(Q_{d_L}+Q_{d_R}),
\end{equation}
such that the effective coupling of the DM particle to protons and neutrons is,
\begin{equation}
f_p=\frac{g_{dm}g'}{2m_{Z'}^2}(2Q_{u_L}+2Q_{u_R} + Q_{d_L}+Q_{d_R}),
\end{equation}
\begin{equation}
f_n=\frac{g_{dm}g'}{2m_{Z'}^2}(Q_{u_L}+Q_{u_R} + 2Q_{d_L}+2Q_{d_R}),
\end{equation}
We can define the amount the isospin violation as the ratio $f_n/f_p$, that in our case is given by
\begin{equation}
f_n/f_p=\frac{Q_{u_L}+Q_{u_R} + 2Q_{d_L}+2Q_{d_R}}{2Q_{u_L}+2Q_{u_R} + Q_{d_L}+Q_{d_R}}.
\end{equation}
As we can see, the ratio $f_n/f_p$,
depends \emph{exclusively} on the charges of the corresponding quarks under the extra $U(1)_X$ symmetry. Having a non-Universal extra gauge symmetry means that in typical realisations these charges will not be the same, and therefore the amount of isospin violation will in general be different than the usual $\pm 1$ of Universal models, thus providing a distinctive, particular signal in the cross section of DM-nucleus elastic scattering experiments. Moreover, notice that the vector coupling ratio is independent of the value of the gauge coupling $g'$.

The striking feature of the non-Universal U(1)$\mu\nu$SSM is that the amount of isospin violation can only acquire a discrete set of values. In the non-Universal U(1)$\mu\nu$SSM there is a class of solutions for which both the up and down quarks have the same kind of mass terms (i.e, either tree-level or non-holomorphic). This means in 
particular that it is the opposite Higgs, namely $H_1$ or $H_2$ which gives the mass to each of them. As each Higgs has a different extra charge under the new gauge symmetry, for this class 
of models the amount of isospin violation is parametrised by the Higgs charges as follows
\be
\frac{f_n}{f_p} = \frac{Q_{H_2} + 2Q_{H_1}}{2Q_{H_2} + Q_{H_1}}.
\ee

Hence, as long as the Higgs extra charges are different, a condition necessary in order to forbid the $\mu$ term from the superpotential, there will be isospin violation. It is important to note that the ratio $f_n/f_p$ depends \emph{only} on the Higgs charges and not those of the up and down quark charges.  For example, the model presented in the Appendix~\ref{sec:app} has isospin violation $\approx$ -1.75. Of utmost importance is to stress that not just any value of isospin violation is allowed but, as the extra charges must fulfil certain conditions, the number of possible models is constrained, allowing experiments to discriminate among realisations of the supersymmetric model, which could bear direct relation with the kind of low energy scale string realisation.

%CONCLUSIONS

\section{Conclusions}

In this letter, we have presented a new supersymmetric model in which, by adding a non-Universal U(1) gauge symmetry to the already explored $\mu\nu$SSM model, not only both the $\mu$ term problem and the neutrino masses problem are solved, but the stability of the proton is ensured by recovering an effective R-parity, forbidding at the same time baryon number violating operators and avoiding a possible domain wall problem. By allowing non-Universal charges in all the fields, we demonstrate that there exist families of solutions which require of no exotic matter whatsoever to cancel anomalies. Moreover, in solving the anomalies equations we find that there only is a discrete set of possible extra charges allowed, a fact that has deep implications in the possible phenomenology of the model. 

Universality breaking implies, in the U(1)$\mu\nu$SSM, that the SM fermions will ordinarily have different values of the charge $Q_X$ both for each family and also between up and down doublet 
components. With the direct consequence that, while stringent bounds are already imposed in the possible production of a $Z'$ at accelerators, these bounds do not apply directly to our model. As the 
production rates are calculated assuming that all fermions will couple with the same strength to the $Z'$ boson, and such coupling depends on the specific fermion extra charge, they have to be 
recalculated for each specific realisation of anomalies cancellation in the non-Universal U(1)$\mu\nu$SSM, meaning that new scenarios open up in which even a light $Z'$ boson could have escaped 
detection at LHC. Such scenarios would be constrained only by the condition that no FCNC are observed in the model. A forthcoming work will address in detail the specifics of such scenarios 
\cite{nextone}.

The extra charges dependence of fermionic coupling to a $Z'$ also modifies interaction rates with a DM particle, which is dominated by the lightest $Z'$ mass eigenstate. The conditions imposed by anomalies cancellation lead to a family of scenarios where isospin violation is realised and depends \emph{solely} on the Higgs extra charges, a particle in principle completely unrelated with DM interactions. The implications for DM detection are profound, as the specific model realisation implies a very specific interaction rate with protons and/or neutrons, rendering particular experiments more or less suitable, and modifying the conditions for existing ones. Benchmark scenarios with concrete realisations of extra charge distributions will be analysed in \cite{nextone}.

Summarising, the non-Universal U(1)$\mu\nu$SSM is an attractive model, which could be easily related with specific intersecting brane constructions and which, through a very particular phenomenology consequence of the discrete and constrained extra charge values, could when and if SUSY is discovered, be related directly with the kind of low energy stringy realisation.

%\newpage
\section*{Acknowledgements}

We thank Carlos Mu\~{n}oz and Daniel E. L\'opez-Fogliani for useful discussions and comments on the manuscript. V.M.L. acknowledges support of the Consolider MULTIDARK
project CSD2009-00064, the SPLE ERC project
and the BMBF under project 05H15PDCAA. S.O.C. acknowledges support from MINECO FEDER funds FIS2015-69512-R and Fundaci{\'o}n S{\'e}neca (Murcia, Spain) Project No. ENE2016-79282-C5-5-R.

%APPENDIX

\clearpage
\appendix
\section{Example of charge assignation}\label{sec:app}

We present here two examples of solutions for the anomalies equations, representative of the two main families of models presented in the main text.

\vspace*{0.5cm}

\noindent\textbf{Example 1}

\vspace*{0.5cm}

A minimalist charge assignment that fulfils the anomalies equations and permits the effective $\mu$ term, having a quarks fields hierarchy, and with tree-level Yukawas and non-holomorphic terms appearing at opposing families in quarks. Particularly, in the example presented is the second family of up-quarks and the first, and third families of down quarks, the ones which acquire mass via non-holomorphic terms.

\begin{center}
%\label{Table3}
\begin{tabular}{||c | c | c| c| c| c| c||}
\hline
\hline
 & $Q_Q$ & $Q_u$ & $Q_d$ & $Q_L$ & $Q_e$ & $Q_{\nu^c}$ \\
\hline
$1^{st}$ Family & -$\frac{1}{3}$ & $\frac{-2}{3}$  & $\frac{4}{3}$  & 2 & -4 & -3 \\
\hline
$2^{nd}$ Family & -$\frac{5}{3}$ & $\frac{11}{3}$ & -$\frac{1}{3}$ & 7 & -6 & -8 \\ 
\hline
$3^{rd}$ Family & -$\frac{8}{3}$ & $\frac{5}{3}$  & $\frac{11}{3}$  & 2 & -4 & -3 \\
\hline
\multicolumn{3}{||c|}{ $Q_{H_1}$ = 2 } & \multicolumn{4}{|c||}{ $Q_{H_2}$ = 1}  \\
\hline
\hline
\end{tabular} 
\end{center}

With this charges assignment, the fields mass terms in the superpotential below the soft breaking scale read
\be
\begin{split}
-\mathcal{L}_{eff} &= 
Y_u(u_L)^c q_u H_2 + \tilde{Y}_c(c_L)^c q_c H_1 + Y_t(t_L)^c q_t H_2 \\&
+ \tilde{Y}_d(d_L)^c q_d H_2 + Y_s(s_L)^c q_s H_1 + \tilde{Y}_b(b_L)^c q_b H_2 \\&
+ Y_e(e_L)^c L_e H_1 + \tilde{Y}_\mu(\mu_L)^c L_\mu H_2 + Y_\tau(\tau_L)^c L_\tau H_1 \\&
+ \epsilon_{ab} Y_\nu^{ij} \,  H_2^b\,  L^a_i \,  \nu^c_j + \nu^c_e\,\hat H_1 \hat H_2 + + \nu^c_\tau\,\hat H_1 \hat H_2,
\end{split} 
\label{superpotential2}
\ee
where the tilded Y represent the Yukawas generated by non-holomorphic interactions. 

\vspace*{0.5cm}

\noindent\textbf{Example 2}

\vspace*{0.5cm}

Now we present an example of solution for the family of models which have non-trivial violation of isospin, with consequences for the phenomenology of dark matter. In this case, the condition is that both first families of up and down quarks have the same kind of mass term. In this case, having both with a Yukawa at tree-level, a possible anomalies cancellation charges distribution is
\begin{center}
%\label{Table3}
\begin{tabular}{||c | c | c| c| c| c| c||}
\hline
\hline
 & $Q_Q$ & $Q_u$ & $Q_d$ & $Q_L$ & $Q_e$ & $Q_{\nu^c}$ \\
\hline
$1^{st}$ Family & $\frac{1}{3}$ & $\frac{4}{3}$  & $\frac{-7}{3}$  & -10 & $\frac{25}{3}$ & $\frac{35}{3}$ \\
\hline
$2^{nd}$ Family & -$\frac{-5}{3}$ & $\frac{11}{3}$ & 0 & $\frac{8}{3}$ & -$\frac{14}{3}$ & -1 \\ 
\hline
$3^{rd}$ Family & $\frac{9}{3}$ & -$\frac{4}{3}$  & -$\frac{14}{3}$  & 2 & -4 & -$\frac{1}{3}$ \\
\hline
\multicolumn{3}{||c|}{ $Q_{H_1}$ = 2 } & \multicolumn{4}{|c||}{ $Q_{H_2}$ = -$\frac{5}{3}$}  \\
\hline
\hline
\end{tabular} 
\end{center}

\bibliography{biblio.bib}
\bibliographystyle{JHEP}

\end{document}